\documentclass[]{hdsr}

\graphicspath{{figs/}}

\begin{document}

\newgeometry{bottom=1.5in}


\begin{center}
\LARGE{\textbf{nlive: an R Package to facilitate the application of the sigmoidal and random changepoint mixed models}}
  \title{} 
  \maketitle

  \thispagestyle{empty}

  \begin{tabular}{cc}
  \normalsize{
    Ana W. Capuano\upstairs{\affilone} and
    Maude Wagner\upstairs{\affilone}} 
    \\ [0.25ex]
   \small{\affilone~Rush Alzheimers Disease Center, Rush University Medical Center, Chicago, IL, USA} 
  \end{tabular}
  
  \emails{~~~~~~~~~~~~      Corresponding author: ana\_capuano@rush.edu}
  \vspace*{0.1in}

\begin{abstract}

\noindent \textbf{Background}: The use of mixed effect models with a specific functional form such as the Sigmoidal Mixed Model and the Piecewise Mixed Model (or Changepoint Mixed Model) with abrupt or smooth random change allows the interpretation of the defined parameters to understand longitudinal trajectories. Currently, there are no interface R packages that can easily fit the Sigmoidal Mixed Model allowing the inclusion of covariates or incorporating recent developments to fit the Piecewise Mixed Model with random change. \noindent \textbf{Results}: To facilitate the modeling of the Sigmoidal Mixed Model, and Piecewise Mixed Model with abrupt or smooth random change, we have created an R package called nlive. All needed pieces such as functions, covariance matrices, and initials generation were programmed. The package was implemented with recent developments such as the polynomial smooth transition of the piecewise mixed model with improved properties over Bacon-Watts, and the stochastic approximation expectation-maximization (SAEM) for efficient estimation. It was designed to help interpretation of the output by providing features such as annotated output, warnings, and graphs. Functionality, including time and convergence, was tested using simulations. We provided a data example to illustrate the package use and output features and interpretation. The package implemented in the R software is available from the Comprehensive R Archive Network (CRAN) at https://CRAN.R-project.org/package=nlive. \noindent \textbf{Conclusions}: The nlive package for R fits the Sigmoidal Mixed Model and the Piecewise Mixed: abrupt and smooth. The nlive allows fitting these models with only five mandatory arguments that are intuitive enough to the less sophisticated users.
\end{abstract}
\end{center}

  \small	

\section{BACKGROUND}
\noindent
Continuous longitudinal data may have a trajectory that is not linear. This is the case of cognitive aging and other processes in fields such as agriculture [1], pharmacology [2], and marketing [3]. Although some less parsimonious models have been proposed to model such longitudinal data, the use of models with a specific functional form such as the Sigmoidal Mixed Model (SMM) [4] and the Piecewise Mixed Model (PMM) [5] with abrupt or smooth change allow the interpretation of the defined parameters.
\\
The SMM is currently implemented in SAS using PROC NLMIXED [4], which maximizes the marginal likelihood by using an adaptive Gaussian quadrature [6] or other approximation methods, such as the first-order method [7]. In R, most of these packages focus on dose-response optimization and curve-fitting  [8] such as $qpcR$ [9], $grofit$  [10], $FlexParamCurve$  [11], $drfit$  [12], and $MCPMod$  [13] or aim to automate fitting and classify multiple curves  [8]. However, none of these packages can fit the SMM allowing the inclusion of covariates for all 4 parameters. The PMM is commonly fitted using Bayesian inference and implemented in OpenBugs or WinBUGS, but is also commonly fit in R using the lme4  [14], which maximizes the marginal likelihood by using a Laplace approximation. A recently developed Stochastic Approximation Expectation Maximization (SAEM) algorithm was shown to be more successful  [15] and faster  [16] to identify the maximum likelihood estimators of non-linear mixed models. This can be implemented directly using the package saemix  [17] (version 3.0). However, one downside of having such flexible packages as the lme4 and saemix is that they require more analytical skills to code. It is worth noting that some simple-to-use packages in R can fit the abrupt PMM, including segmented  [18] and rcpm  [19]. However, these packages also do not use SAEM and with them, it is not possible to (i) include covariates for all 4 parameters, (ii) consider a smooth polynomial transition, and/or (iii) estimate directly the last level (e.g. level close to death).
\\
In this work, we present the the version 0.2.0 of the $nlive$ package implemented within R software. The main objective of the package is to facilitate and broaden the application and interpretation of the SMM and PMM for longitudinal data. All needed elements to fit the models have been programmed, including the computation of the structural model and the automatic generation of initials for the main parameters. As such, less experienced R users only need to specify the model to fit via a single intuitive line of code, with only five mandatory arguments. The package was implemented with the most recent and efficient algorithms for non-linear models. Implementation was also performed with the most interpretable parameterization and was based on the most recent developments in each type of model. For example, for the smooth PMM, instead of using the Bacon–Watts  [20] which can create an artificial increase in the trajectory right after the changepoint  [21], we considered the most recently developed polynomial smooth transition [22]. In this article, we reintroduce these models, describe the implementation of the package, and provide a simulation study to demonstrate the performance of the package. We also demonstrate the use of the model and interpretation of the output using a  made-up dataset with trajectories similar to those found in the cognitive aging of individuals followed until death  [23].

\section{MODEL SPECIFICATIONS}
\noindent
As a prelude to the introduction and demonstration of the new $nlive$ package, we first describe the general formulation of the nonlinear mixed models implemented in the package. The simplified general form of nonlinear mixed models can be written in terms of a known nonlinear function $f$ given by:
\begin{equation}\label{1}
~~~~~~~~~~~~~~~~~~~~~~~~~~~~~~~~~y_{ij}=f(t_{ij},\psi_i)+\epsilon_{ij}
\end{equation}

\noindent where $y_{ij}$ denotes the longitudinal outcome value of subject $i$ $(i=1, ..., N)$ collected at the observation time $t_{ij}$ $(j=1, ..., n_i)$; $\psi_i$ is a vector of normally distributed person-specific parameters function of fixed effects and individual random effects; and $\epsilon_{ij}$ are random error, with $\epsilon_{ij} \sim N(0, \sigma_{\epsilon}^2)$. 
\\
Motivated by the application on late-life cognitive decline, the  $nlive$ package implements two main classes of nonlinear mixed models: the Sigmoidal Mixed Model (SMM)  [4,24] with four parameters and the Piecewise Linear Mixed Model (PMM) [5] with two linear phases and a single changepoint. In the following sub-sections, we provide a brief introduction to these models. For simplicity, some annotations can be similar from one model to another, while the interpretation of the parameters remains specific to each of them.
 
\subsection{The Sigmoidal Mixed Model}
The SMM introduced by Capuano and colleagues [4] is based on the four-parameter logistic that allows the inclusion of covariates related to four parametric quantities. The non-linear trajectory of the outcome $Y$ can be formulated as follows: 

\begin{equation}
\label{2}
~~~~~~~~~~~~~~~~~~~~~~~~f(t_{ij}, \psi_i) = \psi_{1i} + \frac{\psi_{2i}  - \psi_{1i}}{1+(t_{ij}/\psi_3)^{\psi_4}} 
\end{equation}
\\
\noindent where the first parameter, $\psi_{1i}$, represents the person-specific initial level of the outcome before the onset of decline. The second parameter, $\psi_{2i}$, represents the person-specific level of the outcome at a time equal to zero (e.g., death), or the intercept. We will call it the last level although the meaning of time may differ depending on the application. $\psi_{3}$ represents the marginal time when half of the total decline occurred. We will call it the midpoint. $\psi_{4}$ represents the marginal Hill slope and will define the nonlinear pattern of the trajectory (e.g. determining the steepness, earlier versus later acceleration of change). These two latter parameters are kept as marginal for convergence purposes [4]. 
The four main parameters are assumed to obey the following equations:
\begin{equation}
\label{3}
~~~~~~~~~\mbox{initial level:}~\psi_{1i} = \alpha_1 + X_{1i}^\top\beta_1 + \eta_{1i}
\end{equation}
\begin{equation}
\label{4}
~~~~~~~~~\mbox{last level (intercept):}~\psi_{2i} = \alpha_2 + X_{2i}^\top\beta_2 + \eta_{2i}
\end{equation}
\begin{equation}
\label{5}
~~~~~~~~~\mbox{midpoint or time of half decline:}~\psi_{3} = \alpha_3 + X_{3i}^\top\beta_3
\end{equation}
\begin{equation}
\label{6}
~~~~~~~~~\mbox{Hill slope:}~\psi_{4} = \alpha_4 + X_{4i}^\top\beta_4
\end{equation}
\\
\noindent where $\alpha_1$, $\alpha_2$, $\alpha_3$, and $\alpha_4$ are the mean values for the last level, initial level, midpoint, and Hill slope, respectively; $X_{1i}$, $X_{2i}$, $X_{3i}$, and $X_{4i}$ are vectors of covariates associated with the vector of fixed effects $\beta_1$, $\beta_2$, $\beta_3$, and $\beta_4$, respectively; and $\eta_{1i}$ and $\eta_{2i}$ are random effects with $(\eta_{1i}, \eta_{2i})^\top \sim MVN(0, B)$ and $B$ assuming correlations between $\eta_{1i}$ and $\eta_{2i}$.

\subsection{The Piecewise Linear Mixed Model with a Random Changepoint}
The PMM model [5] assumes that the stochastic process of the longitudinal outcome is characterized by two or more different phases. Under this class of models, the $nlive$ package implements two PMM models with an abrupt change (PMM-abrupt) [25] and a smooth polynomial transition (PMM-smooth) [22] between the two linear phases. These models provide an appealing statistical approach to detecting the time when the onset of accelerated decline occurs.

\subsubsection{PMM with abrupt change}
The PMM-abrupt model (also known as the linear-linear or the broken-stick mixed model), consists of an intercept at time zero, a slope close to the intercept, a change point at which the slope changes, and a slope after this change point. The non-linear trajectory of the outcome $Y$ can be formulated as follows:

\begin{equation}\label{7}
  ~~~~~~f(t_{ij},\psi_i)=
  \left\{
    \begin{array}{l}
     \psi_{1i} + \psi_{2i}\psi_{4i} + \psi_{3i}(t_{ij}-\psi_{4i})  ~~~ \mbox{ if } t_{ij}<\psi_{4i}\\
      \psi_{1i} + \psi_{2i}t_{ij} ~~~~~~~~~~~~~~~~~~~~~~~~~ \mbox{ if }t_{ij}\geq\psi_{4i}
    \end{array}
  \right.
\end{equation}
\\
\noindent where the first parameter, $\psi_{1i}$, represents the person-specific level of the outcome at time zero, or the intercept; $\psi_{2i}$ represents the person-specific slope before the changepoint; $\psi_{3i}$ represents the person-specific slope after the changepoint; and $\psi_{4i}$ represents the person-specific changepoint time parameter. 
\\
Assuming an alignment at death for example (for interpretation purposes), the parameters $\psi_{1i}$ to $\psi_{4i}$ are supposed to obey the following equations:
\vspace{-3mm}
\begin{equation}
\label{8}
~~~~~~~~~\mbox{last level (intercept)}: \psi_{1i} = \alpha_1 + X_{1i}^\top\beta_1 + \eta_{1i},
\end{equation}
\begin{equation}
\label{9}
~~~~~~~~~\mbox{slope before the changepoint}: \psi_{2i} = \alpha_2 + X_{2i}^\top\beta_2 + \eta_{2i},
\end{equation}
\begin{equation}
\label{10}
~~~~~~~~~\mbox{slope after the changepoint}: \psi_{3i} = \alpha_3 + X_{3i}^\top\beta_3 + \eta_{3i},
\end{equation}
\begin{equation}
\label{11}
~~~~~~~~~\mbox{changepoint time}: \psi_{4i} = \alpha_4 + X_{4i}^\top\beta_4 + \eta_{4i}
\end{equation}
\\
where $\alpha_1$, $\alpha_2$, $\alpha_3$, and $\alpha_4$ are the mean values for the last level, the slope before the change point, the slope after the changepoint, and the changepoint time, respectively; $X_{1i}$, $X_{2i}$, $X_{3i}$, and $X_{4i}$ are vectors of covariates associated with the vector of fixed effects $\beta_1$, $\beta_2$, $\beta_3$, and $\beta_4$, respectively; and $\eta_{1i}$ to $\eta_{4i}$ are random effects with $(\eta_{1i}, \eta_{2i}, \eta_{3i}, \eta_{4i})^\top \sim MVN(0, B)$ and $B$ assuming correlations only between $\eta_{2i}$ and $\eta_{3i}$. 

\subsubsection{PMM with smooth polynomial transition}
The PMM-smooth model is an extension of the PMM-abrupt. The initial smooth PMM proposed by Bacon and Watts [20] includes a hyperbolic tangent transition. In this work, however, we consider a more recent development that considers a smooth polynomial transition introduced in Van den Hout, Muniz-Terrera, and Matthews [22]. In contrast to the PMM-abrupt, the changepoint of the PMM-smooth represents the beginning of a smooth transition.  
\\
In PMM-smooth, the transition is modeled using a third-degree polynomial function fitted between the two straight lines. In the original work [22,26], the intercept parameter cannot be interpreted directly as it reflects the level parameter projection using the early slope at time zero. To allow direct interpretation of the intercept, we re-formulated the PMM-smooth model as: 

\begin{equation}\label{12}
  f(t_{ij},\psi_i) =
  \left\{
    \begin{array}{l}
     \psi_{1i} + \psi_{2i}t_{ij} + (\psi_{3i}-\psi_{2i})(t_{ij}-{\psi_{4i}}+\frac{v}{2}) ~~~~ \mbox{ if } t_{ij}<{\psi}_{4i}\\
     g_{transition} (t_{ij} \mbox{\textbar} \psi_{1i},\psi_{2i},\psi_{3i},v) ~~~~~~~~~~~~~~~~~~~ \mbox{ if } {\psi}_{4i}\leq t_{ij}\leq {\psi}_{4i}+v\\
      \lambda_{i}+\psi_{2i}t_{ij} ~~~~~~~~~~~~~~~~~~~~~~~~~~~~~~~~~~~~~~~~~~ \mbox{ if }t_{ij}>{\psi}_{4i}+v
    \end{array}
  \right.
\end{equation}
where $\psi_{1i}$, $\psi_{2i}$, and $\psi_{3i}$ have been previously defined for Equation \eqref{7}. ${\psi}_{4i}$ is the person-specific time when the smooth transition phase of length $v$ begins. $v$ is a value representing the time interval where the polynomial curve occurs between $t_{ij}$ = ${\psi}_{4i}$ and $t_{ij}$ = ${\psi}_{4i}$ + $v$. To be closer to the PMM-abrupt, the two linear parts should intersect at the middle of the transition phase and the constraint $\lambda_i=\psi_{1i}+\psi_{2i}({\psi_{4i}}+\frac{v}{2})-\psi_{3i} ({\psi}_{4i}+\frac{v}{2})$ is imposed. Note that $v$ set to 0 reduces to a PMM-abrupt model.
\\
The smoothness of the transition function involves four linear equations with four parameters:
\begin{equation}\label{13}
~~~~~~~~~~~~~~g_{transition}({\psi}_{4i}) = \lambda_i+\psi_{3i} {\psi}_{4i}
\end{equation}
\begin{equation}\label{14}
~~~~~~~~~~~~~~g_{transition}({\psi}_{4i}+v) = \psi_{1i}+ \psi_{2i} ({\psi}_{4i}+v)
\end{equation}
\begin{equation}\label{15}
~~~~~~~~~~~~~~(\frac{\partial}{\partial{t_{ij}}}g_{transition})({\psi}_{4i}) = \psi_{3i}
\end{equation}
\begin{equation}\label{16}
~~~~~~~~~~~~~~(\frac{\partial}{\partial{t_{ij}}}g_{transition})({\psi}_{4i}+v) = \psi_{2i}  
\end{equation}
\\
where $g_{transition}$ is obtained by solving the system of four linear equations with four unknown parameters. The derivatives of $g_{transition}$ at the times $t_{ij}={\psi}_{4i}$ and $t_{ij}={\psi}_{4i}+v$ are respectively $\psi_{3i}$ and $\psi_{2i}$.   

\section{IMPLEMENTATION}
\subsection{Software}
To facilitate the application and interpretation of the SMM, PMM-abrupt, and PMM-smooth models for a broader audience, who is not necessarily familiar with statistical programming, we developed a user-friendly R package called “nlive” (\textbf{n}on-\textbf{l}inear mixed models with \textbf{i}nitial \textbf{v}alues \textbf{e}stimated) with R version 4.0.3. All needed elements to fit the models, including the definition of the structural model and the generation of initials for the four main parameters, have been programmed so that the user only needs to specify a single intuitive line of code to fit the model. A variety of options can also be specified. Along with the functions to fit individual models, the package also provides a function that displays the longitudinal data. A made-up data frame, under the name {\fontfamily{lmtt}\selectfont dataCog}, is provided with the package. This data represents cognitive patterns previously observed in cognitive aging when participants are followed until death. Real data on cognitive aging with annual follow-up until death from the Religious Order Study and the Memory and Aging Project (ROSMAP) from the Rush Alzheimer’s Disease Center can be obtained under request at https://www.radc.rush.edu. These cohorts are described elsewhere [23]. The version 0.2.0 of the $nlive$ package is freely available via the Comprehensive R Archive Network (CRAN) at https://CRAN.R-project.org/package=nlive. 

\subsection{Estimation}
The SMM, PMM-abrupt, and PMM-smooth previously described were all estimated using the {\fontfamily{lmtt}\selectfont saemix} package (version 3.0) developed by Comets and colleagues [17]. The {\fontfamily{lmtt}\selectfont saemix} package, among other things, requires the definition of the structural model; thus, first, {\fontfamily{lmtt}\selectfont nlive} includes the structures of the models. For SMM, the {\fontfamily{lmtt}\selectfont nlive} algorithm relied on the {\fontfamily{lmtt}\selectfont SSlogis5()} function of the {\fontfamily{lmtt}\selectfont nlraa} package [27] (version 1.2), which initially defines a 5-parameter logistic curve but can be reduced to a 4-parameter logistic when the $5^{th}$ parameter is fixed to 1. For PMM-abrupt and PMM-smooth, the structure of the models are explicitly coded in the {\fontfamily{lmtt}\selectfont nlive} algorithm. In addition to the output information provided by {\fontfamily{lmtt}\selectfont saemix}, the {\fontfamily{lmtt}\selectfont nlive} package also provides p-values for the main terms [28] that would not be available otherwise.

\subsection{The SAEM algorithm}
The computational technique for maximum likelihood estimation implemented in {\fontfamily{lmtt}\selectfont saemix} is the Stochastic Approximation Expectation Maximization (SAEM) algorithm, which is a stochastic approximation version of the standard EM algorithm proposed by Khuhn and Lavielle [29]. The SAEM algorithm showed to be efficient in the context of non-linear mixed models, converging quickly to the maximum likelihood estimators [16] and achieving better performance than linearization-based algorithms [15]. In preliminary testing during the algorithm coding process, in line with the literature, {\fontfamily{lmtt}\selectfont saemix} showed convergence to the adequate solution more often than two main competing software package [17]: {\fontfamily{lmtt}\selectfont nlme} [6] and {\fontfamily{lmtt}\selectfont lme4} [14]. Note that in {\fontfamily{lmtt}\selectfont saemix}, the likelihood can be computed by linearisation or by importance sampling. In the linearization approach, the likelihood of the Gaussian model is estimated from the nonlinear mixed effects model using the approximation proposed by Lindstrom and Bates [30]. In the importance sampling approach, the likelihood is obtained through a Monte-Carlo stochastic integration and does not require a model approximation. More information on these methods is available in the saemix documentation [17].

\subsection{Initial values}
One of the great advantages of the nlive package is that it has an embedded algorithm that examines the data and automatically provides informative initials. Here we briefly describe this algorithm. For SMM, the four main parameters are the last level, first level, midpoint, and Hill slope. An initial for each parameter needs to be provided. For that, we build upon an algorithm previously developed in SAS by Capuano and colleagues [4] (algorithm hosted and accessible at github.com/AWCapuano/sigmoidal). In this work, we expand this algorithm for different data structures (e.g., different time scales). Briefly, first, the algorithm segments the time into the initial (5th percentile) and final (95th percentile). At these periods the initial and final mean levels of the outcome are ascertained. The time of half decline is set to 300 if the curve is nearly linear, and to 2 otherwise. Finally, the Hill slope is set to a high and low value based (0.5 and 1.05). Similarly, for PMM-abrupt and PMM-smooth, estimation of the models requires the specification of four starting values related to the four main parameters: last level, changepoint time, slope before the changepoint, and slope after the changepoint. First, the algorithm obtains the final portion of the time (the 95th percentile of time). Then, for this period of time, the mean level of the outcome is ascertained. To inform the other three parameters (changepoint, pre-slope, post-slope), standard linear mixed models are used. First, the time is segmented into quintiles starting from the initial time to the final time. Then the algorithm approximates where the acceleration of change (i.e., changepoint) occurs by estimating five separate linear mixed models based on the quintiles of time. The changepoint time is defined as the lower bound of the time interval where the fastest slope occurred. Lastly, the early and final slopes are informed by the slope of cognitive decline estimated using a linear mixed model considering the subsets of cognitive measures collected before and after the approximated changepoint, respectively. The linear mixed models are implemented using the hlme function from the lcmm [31] package (version 1.9.5) to fit mixed effect models on segments of the longitudinal data.

\section{OVERVIEW OF THE PACKAGE}
The {\fontfamily{lmtt}\selectfont nlive} package offers three estimation functions ({\fontfamily{lmtt}\selectfont nlive.smm, nlive.pmma, nlive.pmms}) to fit the SMM, PMM-abrupt, PMM-smooth models, respectively, relying on the SAEM algorithm implemented in {\fontfamily{lmtt}\selectfont saemix}. These functions all require to take as input a dataset that provides information on the longitudinal outcome of interest, participant ID, time, and predictors (if any). \medskip 
\\
The call of {\fontfamily{lmtt}\selectfont nlive.smm} is:
\medskip \\
\small
{\fontfamily{lmtt}\selectfont
\noindent nlive.smm(dataset, ID, outcome, time, var.all=NULL, var.last.level=NULL, 
\\var.first.level=NULL, var.midpoint=NULL, var.Hslope=NULL, traj.marg=NULL, 
\\traj.marg.group=NULL, start=NULL)}
\\\\
\normalsize
The first fourth arguments are mandatory: {\fontfamily{lmtt}\selectfont dataset} defines the name of the data frame in the longitudinal format; {\fontfamily{lmtt}\selectfont ID} defines the name of the variable representing the grouping structure specified with " (e.g., {\fontfamily{lmtt}\selectfont "ID"} representing the unique identifier of participants); {\fontfamily{lmtt}\selectfont outcome} corresponds to the name of the time-varying variable representing the longitudinal outcome specified with " (e.g., {\fontfamily{lmtt}\selectfont "cognition"}); {\fontfamily{lmtt}\selectfont time} is the name of the variable representing the timescale specified with " (e.g., {\fontfamily{lmtt}\selectfont "time"}), which can be negative or positive.
\\
All other arguments are optional: {\fontfamily{lmtt}\selectfont var.all} specifies a vector indicating the name of the variable(s) that the four main parameters of the model will be adjusted to (e.g. {\fontfamily{lmtt}\selectfont var.all=c("X1")})({\fontfamily{lmtt}\selectfont NULL} by default); {\fontfamily{lmtt}\selectfont var.last.level, var.first.level, var.midpoint, var.Hslope}, each specifies a vector indicating the name of the variable(s) that the related specific parameter (e.g. {\fontfamily{lmtt}\selectfont midpoint}) can be adjusted to (e.g. {\fontfamily{lmtt}\selectfont var.midpoint=c("X2")})({\fontfamily{lmtt}\selectfont NULL} by default); {\fontfamily{lmtt}\selectfont traj.marg} indicates whether the marginal estimated trajectories should be plotted ({\fontfamily{lmtt}\selectfont TRUE}) or not ({\fontfamily{lmtt}\selectfont FALSE} by default); {\fontfamily{lmtt}\selectfont traj.marg.group} provides the name of the grouping variable listed in one of the arguments {\fontfamily{lmtt}\selectfont var.} to contrast the estimated marginal trajectories between two specific groups ({\fontfamily{lmtt}\selectfont NULL} by default). If the variable is binary, the trajectories are contrasted between the two groups of interest. If the variable is continuous, the $10^{th}$ and $90^{th}$ percentile values will automatically be considered; {\fontfamily{lmtt}\selectfont start} specifies a vector of length 4 to override the specification of the four initial values for the main parameters (initials obtained directly from the data by default).
\medskip 
\\
The calls of {\fontfamily{lmtt}\selectfont nlive.pmma} and {\fontfamily{lmtt}\selectfont nlive.pmms} are:
\medskip \\
\small
{\fontfamily{lmtt}\selectfont
\noindent nlive.pmma(dataset, ID, outcome, time, var.all=NULL, var.last.level=NULL, 
\\var.slope1=NULL, var.slope2=NULL, var.changepoint=NULL, ...)}
\\\\
{\fontfamily{lmtt}\selectfont
\noindent nlive.pmms(dataset, ID, outcome, time, var.all=NULL, var.last.level=NULL, 
\\var.slope1=NULL, var.slope2=NULL, var.changepoint=NULL, ...)}
\\\\
\normalsize
All arguments needed in {\fontfamily{lmtt}\selectfont nlive.pmma} and {\fontfamily{lmtt}\selectfont nlive.pmms} are the same as those previously described  for {\fontfamily{lmtt}\selectfont nlive.smm}. However, because the parameters are different between the models, the PMM-related functions will have different arguments to enter covariates to specific parameters. These arguments are {\fontfamily{lmtt}\selectfont var.last.level}, {\fontfamily{lmtt}\selectfont var.slope1}, {\fontfamily{lmtt}\selectfont var.slope2}, and {\fontfamily{lmtt}\selectfont var.changepoint}. Of note, {\fontfamily{lmtt}\selectfont nlive} has a legacy function that can fit all models. This function is being kept for older users. Details are available in Supplementary Materials (Appendix 1). Information on all the functions available in {\fontfamily{lmtt}\selectfont nlive} is also found in the package documentation at https://CRAN.R-project.org/package=nlive.

\section{RESULTS}
\subsection{Performance}
\noindent We performed a simulation study to evaluate the performance of fitting SMM and PMM using the SAEM algorithm and of assigning informative initials using the nlive algorithm. In the first step of the study, we challenged the SAEM algorithm by running two different scenarios: varying sample sizes (n of individuals of 100, 200, and 500), and varying number of covariates (zero, one, and two covariates per parameter). In the second step we compared the gain of using the nlive algorithm to assign initials by comparing it with the use of naïve initials, that is using zero to all initials. The simulated data mimicked the longitudinal cognitive trajectoriems observed in deceased ROSMAP participants. These cohorts were chosen due to the large number of participants followed annually until death. The cohorts are described in detail elsewhere [17]. The time intervals for each visit of each individual were generated using a uniform distribution of [–2, 2] months. The data generation started randomly from 24 years before death (time=-24) to death (time=0). This allowed the dataset to be more realistic where each individual has a different time from baseline to death, and there is an average of 10 years of follow-up (SD=5). As an inclusion criterion, individuals need at least 4 cognitive observations to enter the model. Each scenario was tested using 100 replications.
\\

In the first step, for each model, convergence was successfully reached in the second phase of the SAEM algorithm in all replications under all scenarios. The run time increased with the sample size and the number of covariates and was longer for SMM (Figure 1). However, in the most complex scenarios of fitting SMM with two covariates on a sample of 500 individuals, the run time was still less than 6 minutes. We assessed the estimation accuracy of the marginal mean cognitive values before death, year by year, using the empirical Mean Squared Errors ($MSE(t)=\frac{1}{100} \sum_{r=1}^{100}(Y(t) - \hat{Y}_{r}(t))^{2}$), where $Y(t)$ and $\hat{Y}_{r}(t)$ represent the underlying true level and the estimated level of cognition, respectively, at year $t$ ($t$=–24,…,0) before death for $r$ replicates ($r$=1,...,100). All the models provided estimates with low bias (MSE ranged from 0.02 to 0.07).
\\
\noindent In the second step, simulations using naïve initials started with the intermediate challenging scenario of one covariate and 500 individuals. Convergence was not a problem with all models successfully reaching a solution in the second phase of the SAEM algorithm. The naïve initials also did not significantly increase the run time (on average, the gain was $<5$ seconds for SMMs and $< 1$ second for PMMs). However, there was a relevant decrease in the quality of the estimates. Using naïve initials, there was up to $30 \%$  increase in the percent bias in the average marginal estimated cognitive trajectories, with up to $42 \%$  increase in the mean square errors. This was enough empirical evidence to convince us to use informative starting values rather than naive ones. Of note, all the models were fitted on a HP ProBook 400 G6 containing an i7-8565U processor and 16 gigabytes of RAM running R version 4.0.3. Together, these profiling results support that the application of the SAEM algorithm to fit SMM, PMM-abrupt, and PMM-smooth is efficient.
\\

\begin{figure}[h]
\centering
\includegraphics[width=12cm]{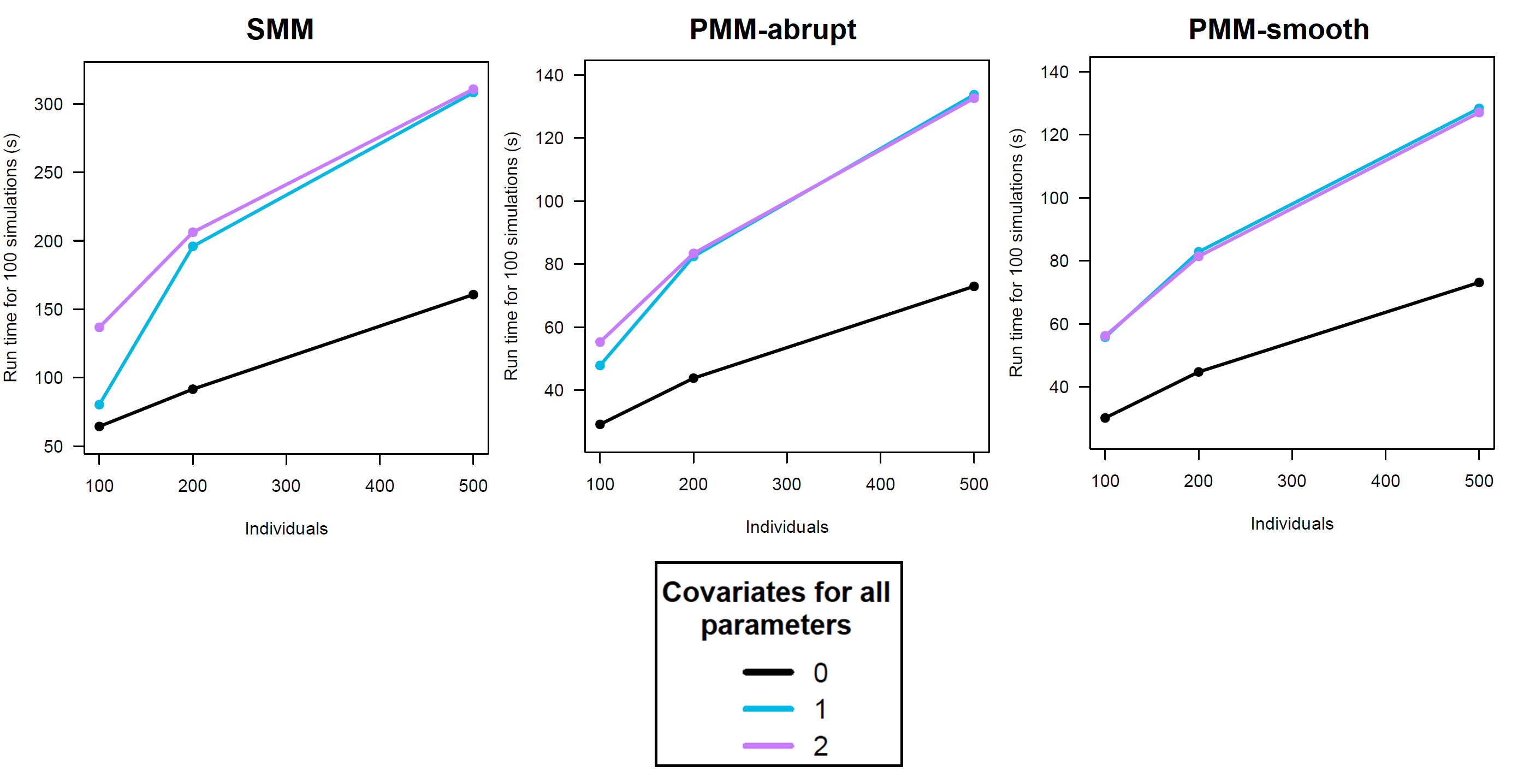}
\caption{Evolution of computation times with the number of individuals, stratified by the number of covariates considered for all the 4 parameters.}
\label{Figure1}
\end{figure}

\newpage
\normalsize
\subsection{Example}
In this section, we show how {\fontfamily{lmtt}\selectfont nlive} can be used to fit the SMM, PMM-abrupt, and PMM-smooth models, and we present the main outputs provided by the package. In the context of our motivating application, late-life cognitive decline, each model was fitted using the made-up illustrative sample {\fontfamily{lmtt}\selectfont dataCog} available in the package. Thus, the first step consists of loading {\fontfamily{lmtt}\selectfont nlive}, which will automatically load {\fontfamily{lmtt}\selectfont dataCog}.
\\\\
\small
{\fontfamily{lmtt}\selectfont 
\noindent R > library(nlive)}
\\\\
\normalsize
The {\fontfamily{lmtt}\selectfont dataCog} dataset contains 1200 individuals with annual cognitive testing for at least 4 years until death (mean follow-up=7 [SD=5] years); a description of the data can be accessed via the command {\fontfamily{lmtt}\selectfont summary(dataCog)}. On each line, we can read the unique participant identifier ({\fontfamily{lmtt}\selectfont ID}), the negative retrospective time before death in years ({\fontfamily{lmtt}\selectfont time}), the repeated values of the composite score of global cognition collected over time ({\fontfamily{lmtt}\selectfont cognition}), and the age at death of individuals in years; in the natural scale ({\fontfamily{lmtt}\selectfont ageDeath}) and centered at its mean ({\fontfamily{lmtt}\selectfont ageDeath90}) for interpretation purposes. The following lines create the continuous {\fontfamily{lmtt}\selectfont ageDeath90} variable and display the first lines of {\fontfamily{lmtt}\selectfont dataCog}: 
\\\\
\small
{\fontfamily{lmtt}\selectfont
\noindent R > dataCog\$ageDeath90 <- dataCog\$ageDeath - 90
\\ R > head(dataCog)   
\\.~~~~ID~~~~time~cognition ageDeath ageDeath90 
\\1~~1000  ~-10.00 ~~~~~0.45 ~~~~~~91 ~~~~~~~~~1
\\2~~1000 ~~-9.08 ~~~~~0.27  ~~~~~~91 ~~~~~~~~~1
\\3~~1000 ~~-8.04 ~~~~~0.19  ~~~~~~91 ~~~~~~~~~1
\\4~~1000 ~~-6.82 ~~~~~0.15  ~~~~~~91 ~~~~~~~~~1
\\5~~1000 ~~-5.99 ~~~~~0.05  ~~~~~~91 ~~~~~~~~~1
\\6~~1000 ~~-4.98 ~~~~~0.15  ~~~~~~91 ~~~~~~~~~1}
\\\\
\normalsize
Before fitting a model, the user can inspect the longitudinal outcome of interest (or other longitudinal variables) using the function  {\fontfamily{lmtt}\selectfont nlive.inspect()}. Rstudio is recommended for this function. Below is an example of the use of {\fontfamily{lmtt}\selectfont nlive.inspect()} for the variable {\fontfamily{lmtt}\selectfont cognition}: 
\\\\
\small
{\fontfamily{lmtt}\selectfont 
\noindent R> nlive.inspect(dataset="dataCog", ID="ID", variable="cognition", time="time")}
\\\\
\normalsize
{\fontfamily{lmtt}\selectfont nlive.inspect()} generates key plots, including the distribution of the longitudinal outcome, a spaghetti plot of the observed individual cognitive trajectories before death for 70 individuals randomly selected in {\fontfamily{lmtt}\selectfont dataCog} (see Figure 2A), and boxplots of the longitudinal observed measures, obtained every year before death, for the whole population (see Figure 2B). Those plots can be customized as needed using the R code provided in Appendix 2. All plots are produced with the {\fontfamily{lmtt}\selectfont ggplot2} package [32] and allow to better appreciate the variability of the trajectories within and between individuals over time. The options {\fontfamily{lmtt}\selectfont plot.xlabel} and {\fontfamily{lmtt}\selectfont plot.ylabel} allows the user to specify a character string to define axes x and y, respectively (e.g. {\fontfamily{lmtt}\selectfont plot.xlabel=c("Years before death")}). For further graphical adjustements, users can access the {\fontfamily{lmtt}\selectfont ggplot} scripts of the plots generated in {\fontfamily{lmtt}\selectfont nlive.inspect} using {\fontfamily{lmtt}\selectfont nlive.inspect} in an R console.  

\newpage

\begin{figure}[h]
\centering
\includegraphics[width=275pt]{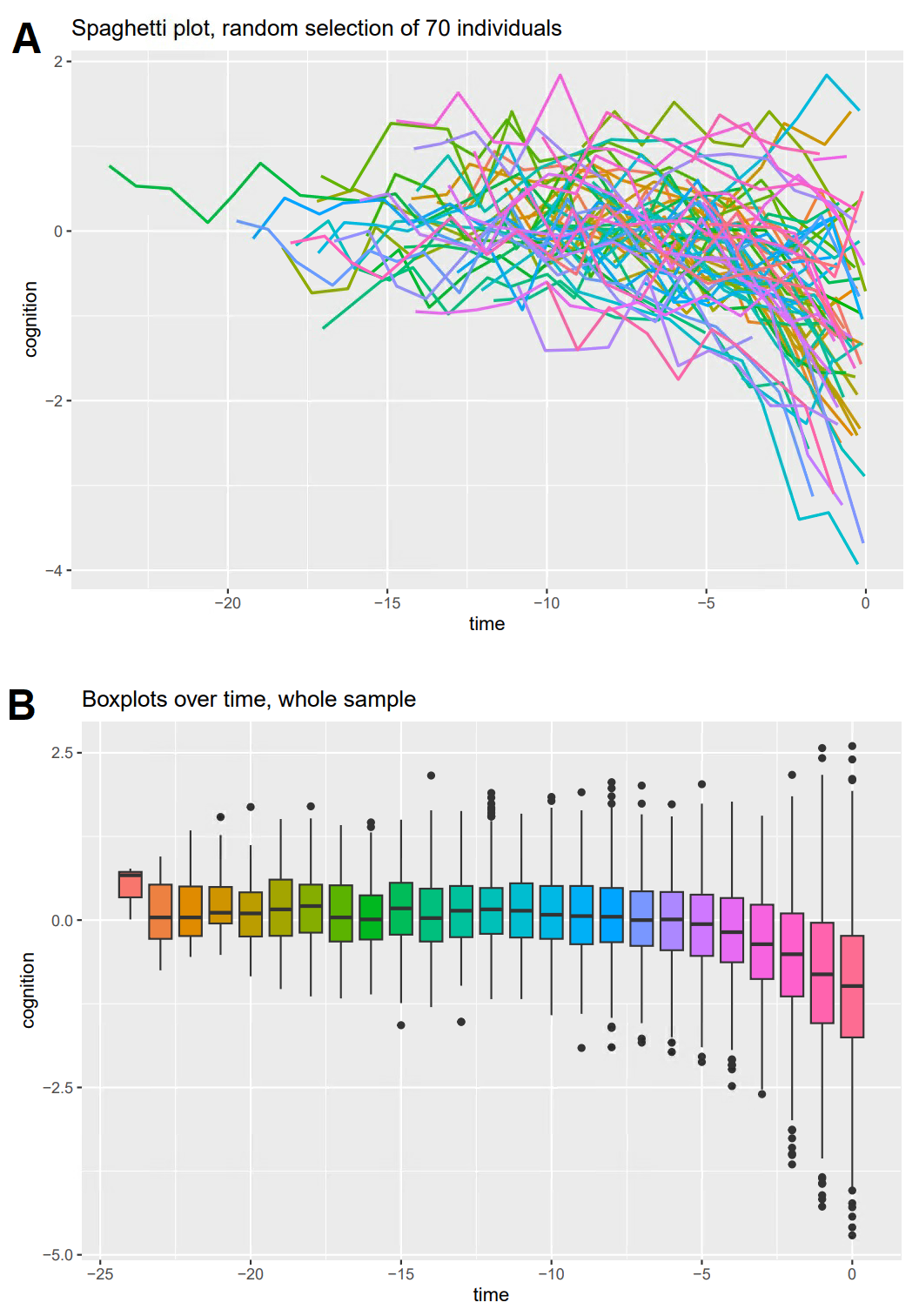}
\caption{\textbf{(A)} Observed individual trajectories of global cognition in the 20 years before death for 70 individuals randomly selected in the made-up illustrative sample {\fontfamily{lmtt}\selectfont dataCog} available in the {\fontfamily{lmtt}\selectfont nlive} package. \textbf{(B)} Boxplots of the longitudinal observed measures, obtained every year before death, in the whole population.}
\label{Figure2}
\end{figure}

\subsubsection{Modeling the SMM}
For demonstration purposes, we fit a relatively simple SMM model with all main parameters adjusted for {\fontfamily{lmtt}\selectfont ageDeath90}. The user only needs to specify the name of the data frame and the columns containing the participant ID, the response, the timescale, and the predictor. We also include arguments to plot the marginal estimated trajectories before death. 
\\\\
\small
{\fontfamily{lmtt}\selectfont
\noindent R > smm.fit <- nlive.smm(dataset=dataCog, ID="ID",
\\+ outcome="cognition", 
\\+ time="time", 
\\+ var.all=c("ageDeath90"), 
\\+ traj.marg=TRUE, 
\\+ traj.marg.group=c("ageDeath90"))}
\\\\
\normalsize
In the main output, \noindent {\fontfamily{lmtt}\selectfont nlive.smm()} provides the general output from {\fontfamily{lmtt}\selectfont saemix}, which include a summary of the data, the specification of the model (main parameters,  covariates, correlation matrix of random effects, initial values), key algorithm options used, and several numerical results (parameter estimated, likelihood) [17]. In \noindent {\fontfamily{lmtt}\selectfont nlive}, this output is augmented by providing the processing time of the program and p-values for the main terms. Below, we focus on the numerical results, but the entire output is displayed in Appendix 3.

\begin{footnotesize}
\begin{lstlisting}
...
----------------------------------------------------
-----------  Variance of random effects  -----------
----------------------------------------------------
            Parameter                  Estimate SE     CV(%)
last.level  omega2.last.level          1.283    0.0556  4.3 
first.level omega2.first.level         0.146    0.0071  4.9 
covar       cov.last.level.first.level 0.049    0.0143 28.9 
----------------------------------------------------
------  Correlation matrix of random effects  ------
----------------------------------------------------
                   omega2.last.level omega2.first.level
omega2.last.level  1.00              0.11              
omega2.first.level 0.11              1.00              
----------------------------------------------------
---------------  Statistical criteria  -------------
----------------------------------------------------
Likelihood computed by linearisation
      -2LL= 9732.152 
      AIC = 9756.152 
      BIC = 9817.233 

Likelihood computed by importance sampling
      -2LL= 9731.84 
      AIC = 9755.84 
      BIC = 9816.921 
----------------------------------------------------
                     Parameter Estimate    SE p-value
1                   last.level   -1.088 0.035 P<.0001
2  beta_ageDeath90(last.level)   -0.061 0.004 P<.0001
3                  first.level     0.24 0.015 P<.0001
4 beta_ageDeath90(first.level)   -0.044 0.002 P<.0001
5                     midpoint   -2.567 0.034 P<.0001
6    beta_ageDeath90(midpoint)    0.031 0.004 P<.0001
7                   hill.slope    1.789  0.04 P<.0001
8  beta_ageDeath90(hill.slope)    0.007 0.005   0.081
9                        error    0.279 0.002 P<.0001
----------------------------------------------------
 The program took 346.51 seconds 
\end{lstlisting}
\end{footnotesize}

\normalsize
\noindent The fitted SMM model indicates that higher age at death was associated with lower cognitive level at baseline (see term {beta\_ageDeath90(first.level)}) and close to death (see term {beta\_ageDeath90(last.level)}). In addition, higher age at death was associated with an earlier half of cognitive decline (see term {beta\_ageDeath90(midpoint)}). However, age at death was not associated with the Hill slope(see term {beta\_ageDeath90(hill.slope)}). 
\\
To facilitate the interpretation of the estimated parameters, it is convenient to visualize the estimated average trajectories over time. In {\fontfamily{lmtt}\selectfont nlive.smm()}, users can easily plot two types of marginal estimated trajectories. First, by setting up the argument {\fontfamily{lmtt}\selectfont traj.marg=T}, the function can provide a graph of the estimated marginal trajectory of global cognition before death in the whole study sample, for the most common profile of covariates (see Figure 3). In this example, this would represent the most common average age at death (i.e., 90 years). Second, by specifying {\fontfamily{lmtt}\selectfont traj.marg.group=c("ageDeath90")}, the function can provide a plot of estimated marginal trajectories of global cognition contrasted between two groups corresponding to participants in the $10^{th}$ versus $90^{th}$ percentile of the {\fontfamily{lmtt}\selectfont ageDeath90} distribution, for the most common profile of covariates (see Figure 4). Users can manually specify the percentile values using the {\fontfamily{lmtt}\selectfont traj.marg.group.val} option. For example, {\fontfamily{lmtt}\selectfont traj.marg.group.val=c(0.25,0.75)} will plot trajectories for the $25^{th}$ and $75^{th}$ percentiles, respectively.

\newpage

\begin{figure}[h]
\begin{center}
\includegraphics[width=230pt]{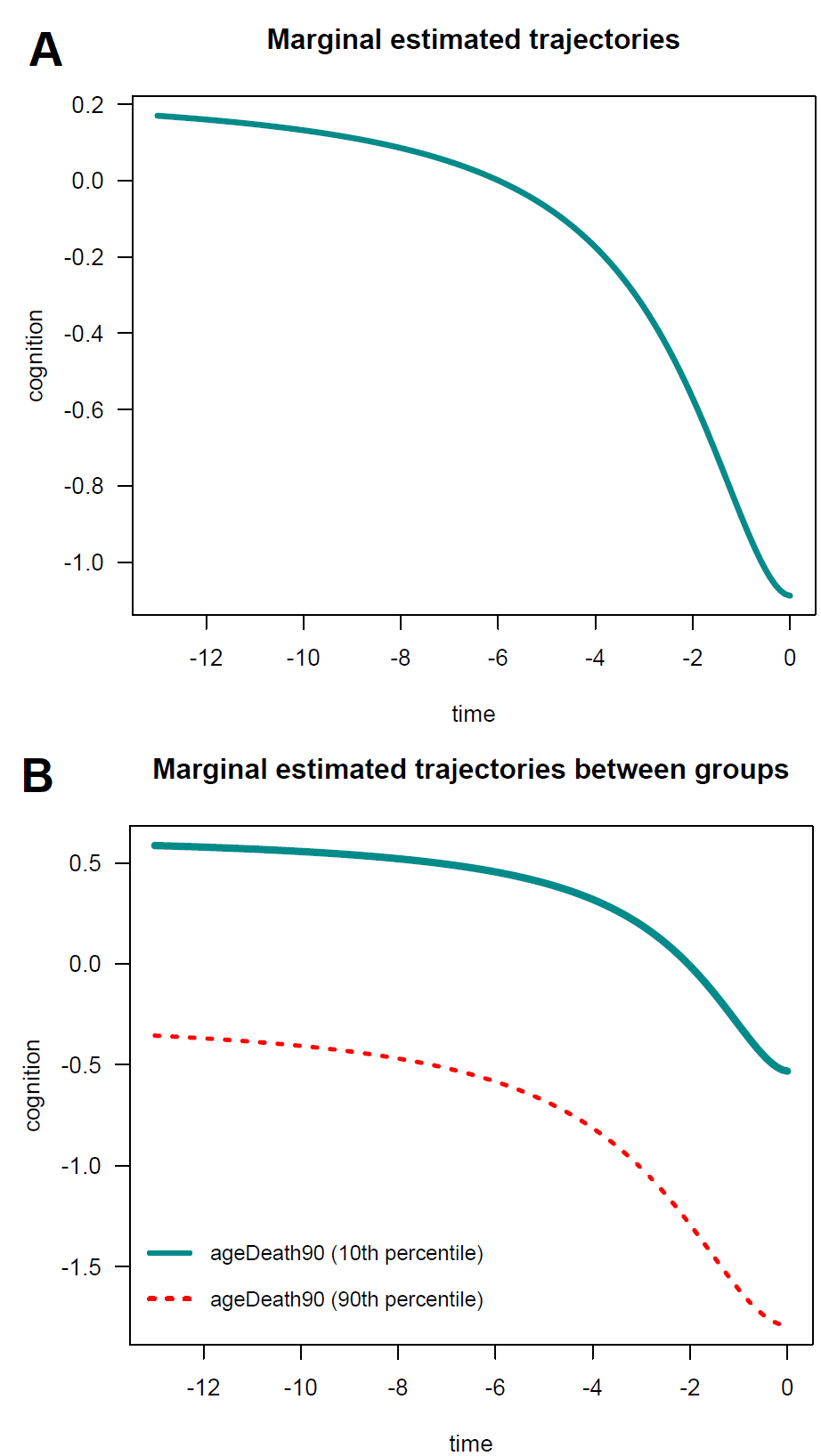}
\caption{Estimated marginal trajectory of global cognition before death \textbf{(A)} for a participant who died aged 90 years (most common profile of covariates), and \textbf{(B)} according to age at death (79 versus 100 years), using the Sigmoidal Mixed Model (n=1,200).}
\end{center}
\label{Figure3}
\end{figure}

\newpage
\normalsize
\subsection{Modeling the abrupt PMM}
For PMM-abrupt, the call of the function {\fontfamily{lmtt}\selectfont nlive.pmma()} is:
\\\\
\small
{\fontfamily{lmtt}\selectfont
\noindent R > pmm.abrupt.fit <- nlive.pmma(dataset=dataCog, ID="ID",
\\+ outcome="cognition", 
\\+ time="time", 
\\+ var.all=c("ageDeath90"), 
\\+ traj.marg=TRUE, 
\\+ traj.marg.group=c("ageDeath90"))}
\\\\
\normalsize
\noindent The general summary output is:

\begin{footnotesize}
\begin{lstlisting}
...      
----------------------------------------------------
-----------  Variance of random effects  -----------
----------------------------------------------------
            Parameter          Estimate SE      CV(%)
last.level  omega2.last.level  1.07196  4.7e-02  4.4 
slope1      omega2.slope1      0.00062  7.4e-05 11.9 
slope2      omega2.slope2      0.03830  2.0e-03  5.2 
changepoint omega2.changepoint 0.58980  7.9e-02 13.4 
covar       cov.slope1.slope2  0.00378  3.2e-04  8.4 
----------------------------------------------------
------  Correlation matrix of random effects  ------
----------------------------------------------------
                   omega2.last.level omega2.slope1 omega2.slope2
omega2.last.level  1                 0.00          0.00         
omega2.slope1      0                 1.00          0.78         
omega2.slope2      0                 0.78          1.00         
omega2.changepoint 0                 0.00          0.00         
                   omega2.changepoint
omega2.last.level  0                 
omega2.slope1      0                 
omega2.slope2      0                 
omega2.changepoint 1                 
----------------------------------------------------
---------------  Statistical criteria  -------------
----------------------------------------------------
Likelihood computed by linearisation
      -2LL= 12349.9 
      AIC = 12377.9 
      BIC = 12449.16 

Likelihood computed by importance sampling
      -2LL= 12296.83 
      AIC = 12324.83 
      BIC = 12396.1 
----------------------------------------------------
                     Parameter Estimate     SE  p-value
1                   last.level   -1.103  0.031  P<.0001
2  beta_ageDeath90(last.level)   -0.062  0.004  P<.0001
3                       slope1   -0.017  0.002  P<.0001
4      beta_ageDeath90(slope1)  -0.0003 0.0004    0.082
5                       slope2   -0.249  0.007  P<.0001
6      beta_ageDeath90(slope2)   -0.001  0.001    0.159
7                  changepoint    -4.25  0.048  P<.0001
8 beta_ageDeath90(changepoint)   -0.059  0.006  P<.0001
9                        error    0.281  0.002  P<.0001
----------------------------------------------------
 The program took 168.58 seconds 
\end{lstlisting}
\end{footnotesize}

\noindent In this example, for the PMM-abrupt model, we found that each additional year of age at death was associated with worse mean cognitive level close to death (see term {beta\_ageDeath90(last.level)}). In addition, each increment in the age at death was related to an earlier onset of accelerated decline (see term {beta\_ageDeath90(changepoint)}). However, age at death was not related to the preterminal decline (see term {beta\_ageDeath90(slope1)}) or terminal decline (see term {beta\_ageDeath90(slope2)}). The marginal estimated trajectories in the whole study sample and in the $10^{th}$ versus $90^{th}$ percentiles of the age at death distribution are displayed in  Figure 4.

\newpage
\begin{figure}[h]
\begin{center}
\includegraphics[width=230pt]{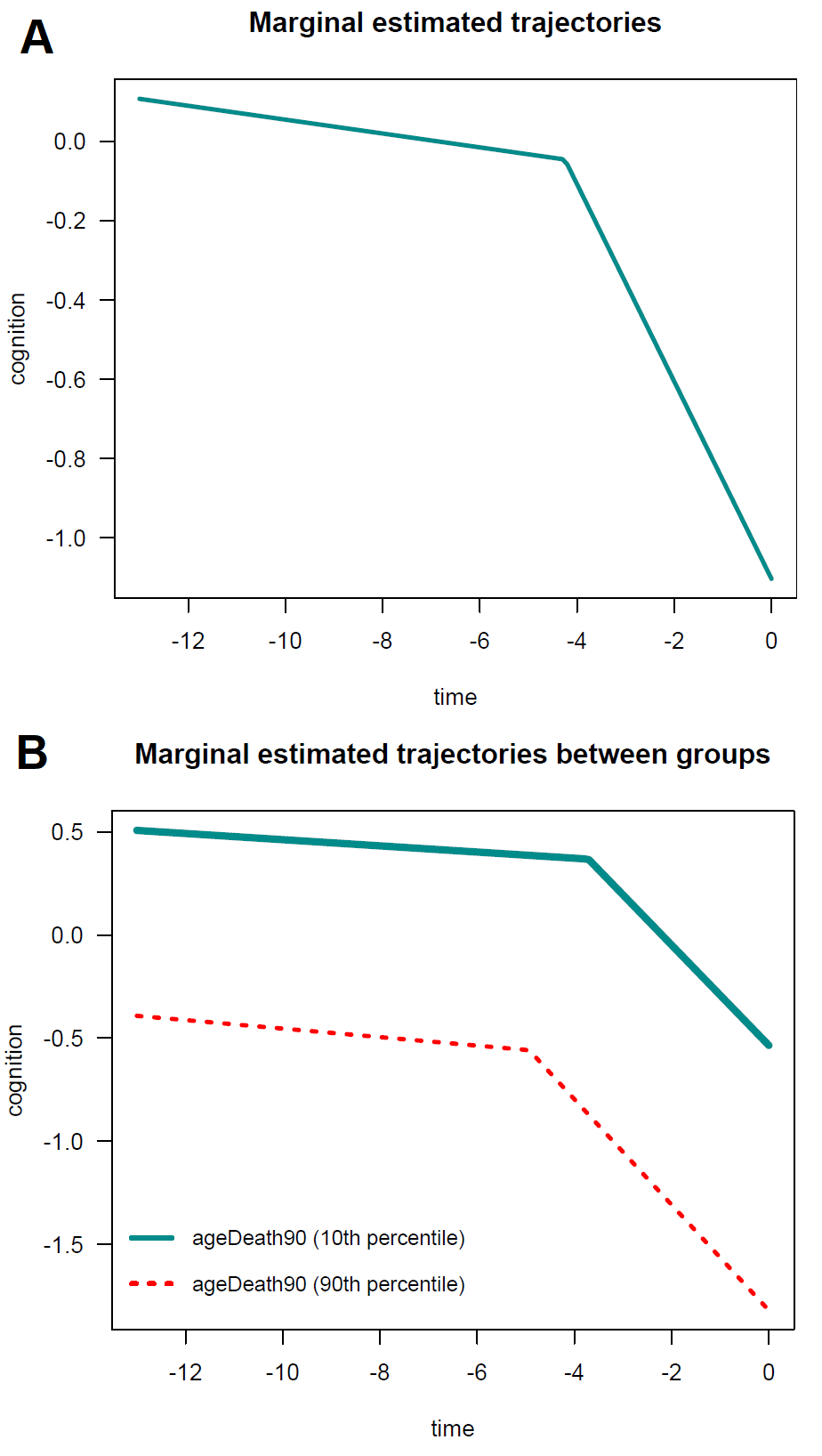}
\caption{Estimated marginal trajectory of global cognition before death \textbf{(A)} for a participant who died aged 90 years (most common profile of covariates), and \textbf{(B)} according to age at death (79 versus 100 years), using the Piecewise Mixed Model with abrupt change (n=1,200).}
\end{center}
\label{Figure4}
\end{figure}

\newpage

\subsection{Modeling the smooth PMM}
\normalsize
For PMM-smooth, the user can call the function {\fontfamily{lmtt}\selectfont nlive.pmms()}:
\\\\
\small
{\fontfamily{lmtt}\selectfont
\noindent R > pmm.smooth.fit <- nlive.pmms(dataset=dataCog, ID="ID",
\\+ outcome="cognition", 
\\+ time="time", 
\\+ var.all=c("ageDeath90"), 
\\+ traj.marg=TRUE, 
\\+ traj.marg.group=c("ageDeath90"))}
\\\\
\normalsize
\noindent The general summary output is:

\begin{footnotesize}
\begin{lstlisting}
...
----------------------------------------------------
-----------  Variance of random effects  -----------
----------------------------------------------------
            Parameter          Estimate SE      CV(%)
last.level  omega2.last.level  1.0699   4.7e-02  4.4 
slope1      omega2.slope1      0.0006   7.3e-05 12.1 
slope2      omega2.slope2      0.0377   2.0e-03  5.2 
changepoint omega2.changepoint 0.6037   8.1e-02 13.3 
covar       cov.slope1.slope2  0.0038   3.1e-04  8.3 
----------------------------------------------------
------  Correlation matrix of random effects  ------
----------------------------------------------------
                   omega2.last.level omega2.slope1 omega2.slope2
omega2.last.level  1                 0.00          0.00         
omega2.slope1      0                 1.00          0.79         
omega2.slope2      0                 0.79          1.00         
omega2.changepoint 0                 0.00          0.00         
                   omega2.changepoint
omega2.last.level  0                 
omega2.slope1      0                 
omega2.slope2      0                 
omega2.changepoint 1                 
----------------------------------------------------
---------------  Statistical criteria  -------------
----------------------------------------------------
Likelihood computed by linearisation
      -2LL= 12357.39 
      AIC = 12385.39 
      BIC = 12456.65 

Likelihood computed by importance sampling
      -2LL= 12293.15 
      AIC = 12321.15 
      BIC = 12392.41 
----------------------------------------------------
                     Parameter Estimate     SE p-value
1                   last.level   -1.099  0.031 P<.0001
2  beta_ageDeath90(last.level)   -0.062  0.004 P<.0001
3                       slope1   -0.017  0.002 P<.0001
4      beta_ageDeath90(slope1)  -0.0003 0.0004   0.082
5                       slope2   -0.246  0.007 P<.0001
6      beta_ageDeath90(slope2)   -0.001  0.001   0.159
7                  changepoint     -5.3  0.049 P<.0001
8 beta_ageDeath90(changepoint)   -0.058  0.006 P<.0001
9                        error    0.281  0.002 P<.0001
----------------------------------------------------
 The program took 134.25 seconds 
\end{lstlisting}
\end{footnotesize}

\noindent In this example, as expected, findings are generally similar to those obtained for the PMM-abrupt model. The main difference is that the estimated changepoint parameter represents here the beginning of the transition period. Marginal estimated trajectories in the whole study sample and according to age at death are displayed in Figure 5, respectively.

\newpage
\begin{figure}[h]
\begin{center}
\includegraphics[width=230pt]{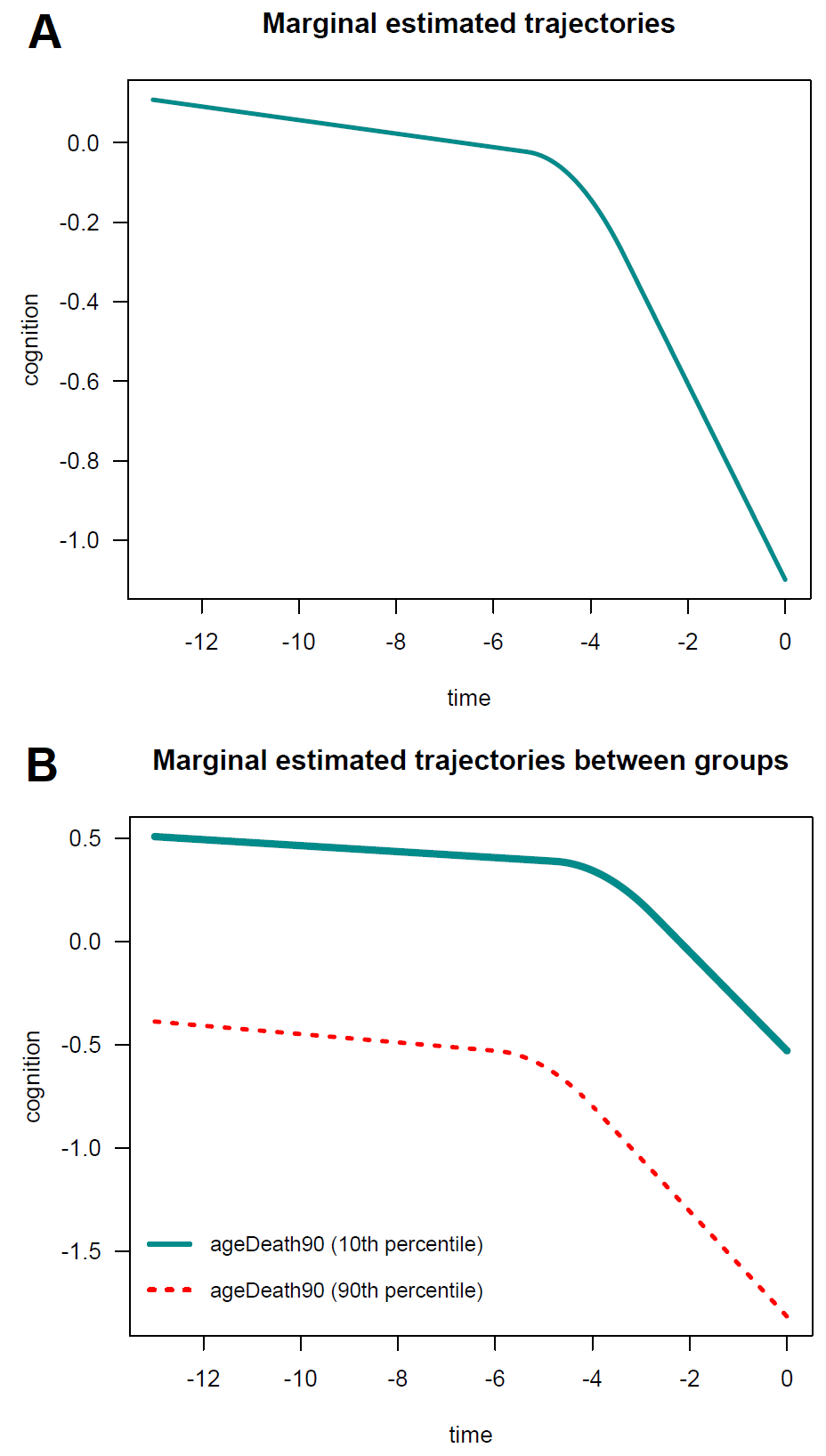}
\caption{Estimated marginal trajectory of global cognition before death \textbf{(A)} for a participant who died aged 90 years (most common profile of covariates), and \textbf{(B)} according to age at death (79 versus 100 years), using the Piecewise Mixed Model with smooth polynomial transition (n=1,200).}
\end{center}
\label{Figure5}
\end{figure}

\section{Other features}
\noindent Since the estimation functions from {\fontfamily{lmtt}\selectfont nlive} fit the models based on the {\fontfamily{lmtt}\selectfont saemix} package, users can take advantage of many generic functions from {\fontfamily{lmtt}\selectfont saemix}. Table 1 displays a brief description of the main functions available in {\fontfamily{lmtt}\selectfont saemix}. Consider the illustrative example of SMM. In this example, {\fontfamily{lmtt}\selectfont nlive} will generate the {\fontfamily{lmtt}\selectfont SaemixObject}  {\fontfamily{lmtt}\selectfont smm.fit}. With that, the user can extract the subject-specific prediction by calling {\fontfamily{lmtt}\selectfont psi(smm.fit, type="mean")}. The subject-specific random effects can be extracted by calling {\fontfamily{lmtt}\selectfont eta(smm.fit, type="mean")}. Convergence plots (Figure 6) can also be obtained by calling {\fontfamily{lmtt}\selectfont saemix.plot.convergence(smm.fit) or} {\fontfamily{lmtt}\selectfont plot(smm.fit, plot.type="convergence")}. Please refer to the {\fontfamily{lmtt}\selectfont saemix} documentation for other available functions. 

\begin{table}[h!]
\caption{Brief description of  functions from {\fontfamily{lmtt}\selectfont saemix} that can be used once model is fit.\\}
    \label{tab:table1}
    \begin{tabular}{p{0.27\linewidth} | p{0.70\linewidth}} 
      \textbf{Function} & \textbf{Description} \\
      \hline     
      {\fontfamily{lmtt}\selectfont summary} & Summary of the data, specification of the model, key algorithm options, and numerical results. 
      \\ {\fontfamily{lmtt}\selectfont plot} & General plot function from SAEM. 
      \\ {\fontfamily{lmtt}\selectfont saemix.plot.fits} & Plot of predictions vs observations for each individual. 
      \\ {\fontfamily{lmtt}\selectfont saemix.plot.obsvspred} & Plot of marginal predictions vs observations, and individual predictions vs observations.
      \\ {\fontfamily{lmtt}\selectfont saemix.plot.convergence} & Plot of parameter estimate vs iteration number for each parameter.
      \\ {\fontfamily{lmtt}\selectfont saemix.plot.randeff} & Boxplot of the random effects. 
      \\ {\fontfamily{lmtt}\selectfont coef} & Vector of the coefficients from a saemix fit. 
      \\ {\fontfamily{lmtt}\selectfont eta} & Subject-specific estimates of the parameters and random effects.
      \\ {\fontfamily{lmtt}\selectfont vcov} & Variance-covariance matrix. 
      \\ {\fontfamily{lmtt}\selectfont logLik} & Likelihood from a saemixObject resulting from a call to saemix.
      \\
      \hline
    \end{tabular}  
\end{table}

\begin{figure}[h]
\centering
\includegraphics[width=355pt]{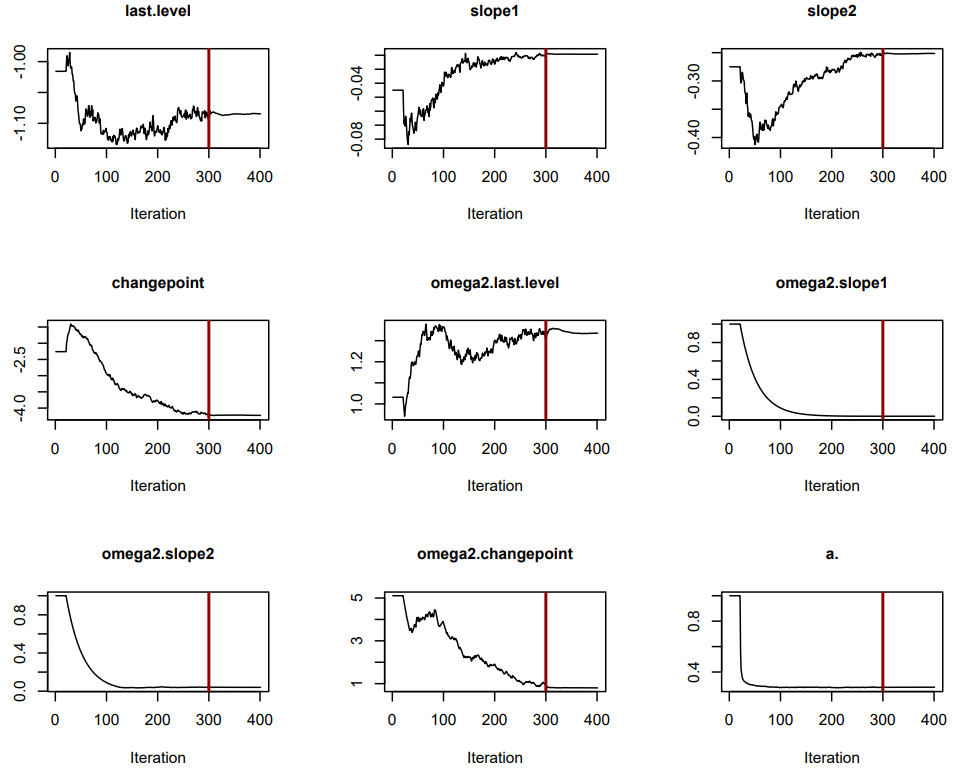}
\caption{Plots evaluating the convergence for the estimation of the parameter.}
\label{Figure6}
\end{figure}

\newpage
\section{CONCLUDING REMARKS}
\noindent 
In this work, we introduce the version 0.2.0 of the newly developed R package {\fontfamily{lmtt}\selectfont nlive} to fit three non-linear mixed models for Gaussian longitudinal data: the sigmoidal mixed model (SMM) and two piecewise linear mixed models with a random changepoint (PMM-abrupt and PMM-smooth). The SMM includes 4 parameters, which allow for the estimation of early level, half of the decline, Hill slope (the steepness of the curve), and final level of the longitudinal outcome of interest. The two PMM separate the trajectory into two linear phases and allow for estimation of the early slope, changepoint, final slope, and final level. These models were chosen for the implementation as they currently cannot be easily implemented in R and are of importance, especially in aging research. All needed pieces such as functions, covariance matrices, and initials generation were programmed. The {\fontfamily{lmtt}\selectfont nlive()} function allows fitting these models with one line of code that is intuitive enough to the less sophisticated users. The yielding product has only five mandatory arguments. Options are available to readily accommodate user preferences, including manual specification of starting values or diagnostic plots. It was also designed to help interpretation of the output by providing features such as annotated output, warnings (e.g. small sample, number of covariates), and graphs.
\\
This package is the first to provide a seamless user interface to fit the Sigmoidal Mixed Effect Model. Some packages in R can fit the Sigmoid curve but not the mixed effect model. As for PMM, although this is not the first package to provide an interface to fit the model, the {\fontfamily{lmtt}\selectfont nlive} package includes the more recent developments in the model structure and the likelihood maximization algorithm. The smooth PMM implemented is based on the polynomial transition that was demonstrated to have improved properties over the Bacon-Watts. PMM models were also reparameterized which allows the interpretation based on the estimated value at time zero and not a projection to zero from the first and more distant slope. Here we build upon recently developed tools in R such as the {\fontfamily{lmtt}\selectfont saemix} package that utilizes the Stochastic Approximation EM-based algorithm, shown in several tests to have a better convergence rate than the Maximum Likelihood. All models were fitted with the same algorithm. Extensive testing of basic functionality was already performed for {\fontfamily{lmtt}\selectfont saemix} development. In this interface, however, convergence adequacy was tested given the particular complexity of these models. Overall the convergence rate was high, the time was reasonable, and the bias was low.  
\\
The motivation of this package was aging research including biomarkers of the Alzheimer’s pathological cascade (a.k.a. Jack curves) [33], natural history of cognition [4,34,35], retesting effect [36,37], and terminal decline [38,39]. These models, however, are non-specific and the {nlive} can be used in a wide variety of fields. Many processes were demonstrated to follow a sigmoid trajectory over time (a.k.a. 3 to 5 parameters logistic, Hill, Langmuir, Langmuir–Hill, and Hill–Langmuir equation). Such processes are found in agriculture [1], pharmacology [2] and marketing [3], to cite a few. Similarly, many processes that are initially linear may have an unknown change that may modify the trajectory. Such processes are found in a wide variety of fields from environmental sciences [40] to engineering [41]. 
\\
In conclusion, we hope that this very user-friendly package will encourage the adoption of more sophisticated models for longitudinal data by the R community, with varying degrees of experience. Although illustrated in the context of cognitive aging, the package can be used in a wide variety of applications.

\section*{Availability and Requirements}
\noindent Project name: nlive R package
\\Project home page: https://cran.r-project.org/web/packages/nlive/index.html
\\Operatimg system: Platform independent
\\Programming language: R
\\Other requirements: No
\\License: MIT 
\\Any restrictions to use by non-academics: No

\section*{Acknowledgments}
The authors thank Dr. Emmanuelle Comets (Université Paris Cité, Université Sorbonne Paris Nord, Inserm, Université Rennes) for kindly reviewing and commenting on several versions of this manuscript and as a collaborator on the programming of the package. The authors also thank Dr. Lisa Barnes (Rush Alzheimer’s Disease Center [RADC]) for reviewing examples during the development of the package using the African American Clinical Core (AA Core) (data not shown), as well as the staff of RADC and the participants of the Minority Aging Research Study (MARS), Religious Orders Study (ROS), and Rush Memory and Aging Project (MAP). Finally, we thank the reviewers for taking the time to go over this article and package and provide important suggestions.

\section*{Abbreviations}
\noindent CRAN, the comprehensive R archive newtwork
\\PMM, piecewise mixed model
\\SAEM, stochastic approximation expectation maximization
\\SMM, sigmoidal mixed model

\section*{Funding}
The study was supported by NIA grants Grants (R01AG 022018, P30AG072975, R01AG17917) and the Illinois Department of Public Health. Dr. Maude Wagner is supported by a post-doctoral fellowship from the French Foundation for Alzheimer’s Research (alzheimer-recherche.org). The funding organizations had no role in the design or conduct of the study; the collection, management, analysis, or interpretation of the data; or the writing of the report or the decision to submit it for publication.

\section*{Availability of data and materials}
The R package {\fontfamily{lmtt}\selectfont nlive} can be installed from CRAN directly using in an R console {\fontfamily{lmtt}\selectfont install.packages("nlive")}. Archived versions are available from the CRAN at https://CRAN.R-project.org/package=nlive. The most recent update of the {\fontfamily{lmtt}\selectfont nlive} package can be installed from Github by running the command {\fontfamily{lmtt}\selectfont remotes::install\_github("MaudeWagner/nlive")}. The made-up illustrative dataset analyzed during the current study is bundled with the R package {\fontfamily{lmtt}\selectfont nlive}, and can be accessed by running the command {\fontfamily{lmtt}\selectfont data(dataCog, package = "nlive")}. The R code to replicate the simulation study can be provided by the corresponding author upon reasonable request.

\section*{Authors' contributions}
AWC was responsible for the study conception and design. AWC and MW were involved in the analysis and interpretation of the data, drafting of the manuscript and critically revising the manuscript. MW performed the simulations. All authors take responsibility for the integrity of the data and the accuracy of the data analysis. All authors read and approved the final manuscript. 

\section*{Competing interests}
The authors declare that they have no competing interests. 

\section*{Consent for publication}
Not applicable.

\section*{Ethics approval and consent to participate}
Not applicable.

\vspace{6mm}
\section*{REFERENCES}
\vspace{6mm}

1. Zub H, Rambaud C, Bethencourt L, Brancourt-Hulmel M. Late emergence and rapid growth maximize the plant development of Miscanthus clones. Psychology and aging. 2012;5(4):841-54.

2. Gesztelyi R, Zsuga J, Kemeny-Beke A, Varga B, Juhasz B, Tosaki A. Hill equation and the origin of quantitative pharmacology. Archive for history of exact sciences. 2012;66(4):427-38.

3. Gesztelyi R, Zsuga J, Kemeny-Beke A, Varga B, Juhasz B, Tosaki A. Strategic Marketing for High Technology Products: An Integrated Approach. Routledge; 2018.

4. Capuano A, Wilson R, Leurgans S, Dawson J, Bennett D, Hedeker D. Sigmoidal mixed models for longitudinal data. Statistical methods in medical research. 2018 Mar;27(3):863-75.

5. Hall C, Lipton R, Sliwinski M, Stewart W. A change point model for estimating the onset of cognitive decline in preclinical Alzheimer’s disease. Statistics in medicine. 2000;19(11-12):1555-66.

6. Pinheiro J, Bates D. Approximations to the log-likelihood function in the nonlinear mixed-effects model. Journal of computational and Graphical Statistics. 1995;4(1):12-35.

7. Beal S, Sheiner L. Heteroscedastic nonlinear regression. Technometrics. 1988;30(3):327-38.

8. Caglar M, Teufel A, Wilke C. Sicegar: R package for sigmoidal and double-sigmoidal curve fitting. PeerJ. 2018;6:e4251. 

9. Ritz C, Spiess A. qpcR: an R package for sigmoidal model selection in quantitative real-time polymerase chain reaction analysis. Bioinforma (Oxford, England). 2008;24(13):1549–51.

10. Kahm M, Hasenbrink G, Lichtenberg-Frate H, Ludwig J, Kschischo M. Grofit: Fitting biological growth curves. Nat Precedings. 2010;ID:Kahm2010. 

11. Oswald S, Nisbet I, Chiaradia A, Arnold J. FlexParamCurve: R package for flexible fitting of nonlinear parametric curves. Methods Ecol Evol. 2012;3(6):1073–7.

12. Ranke J. Fitting dose-response curves from bioassays and toxicity testing. R News. 2006;3:7–12.

13. Bornkamp B, Pinheiro J, Bretz F. MCPMod: An R Package for the Design and Analysis of Dose-Finding Studies. J Stat Softw. 2009;29(7):1. 

14. Bates D, Mächler M, Bolker B, Walker S. Fitting linear mixed-effects models using lme4. arXiv preprint arXiv:14065823. 2014.

15. Plan E, Maloney A, Mentré F, Karlsson M, Bertrand J. Performance comparison of various maximum likelihood nonlinear mixed-effects estimation methods for dose–response models. The AAPS journal. 2012;14(3):420-32.

16. Delyon B, Lavielle M, Moulines E. Convergence of a stochastic approximation version of the EM algorithm. Annals of statistics. 1999:94-128. 

17. Comets E, Lavenu A, Lavielle M. Parameter Estimation in Nonlinear Mixed Effect Models Using saemix, an R Implementation of the SAEM Algorithm. Journal of Statistical Software. 2017 08/01;80. 

18. Muggeo V. Segmented: an R package to fit regression models with broken-line relationships. R news. 2008;8(1):20-5.

19. Segalas C, Amieva H, Jacqmin-Gadda H. A hypothesis testing procedure for random changepoint mixed models. Statistics in medicine. 2019;38(20):3791-803.

20. Bacon D, Watts D. Estimating the transition between two intersecting straight lines. Biometrika. 1971 12/01;6/11;58(3):525-34. 2022.

21. Richardson S, Lawson R, Davis D, Stephan B, Robinson L, Matthews F, et al. Hospitalisation without delirium is not associated with cognitive decline in a population-based sample of older people—results from a nested, longitudinal cohort study. Age and Ageing. 2021;50(5):1675-81.

22. den Hout AV, Muniz-Terrera G, Matthews F. Smooth random change point models. Statistics in medicine. 2011;30(6):599-610.

23. Bennett D, Buchman A, Boyle P, Barnes L, Wilson R, Schneider J. Religious Orders Study and Rush Memory and Aging Project. Journal of Alzheimer’s disease : JAD. 2018;64(s1):S161-89.

24. Gottschalk PG, Dunn JR. The five-parameter logistic: a characterization and comparison with the four-parameter logistic. Analytical Biochemistry. 2005;343(1):54-65.

25. Hinkley D. Inference about the intersection in two-phase regression. Biometrika. 1969;56(3):495-504.

26. Hout AVD, Muniz-Terrera G, Matthews F. Change point models for cognitive tests using semi-parametric maximum likelihood. Computational Statistics \& Data Analysis. 2013;57(1):684-98.

27. Miguez F, Pinheiro J. Package ‘nlraa’: Nonlinear Regression for Agricultural Applications; 2021. 

28. Liquet B, Commenges D. Correction of the P-value after multiple coding of an explanatory variable in logistic regression. Stat Med. 2001;20(19):2815–26.

29. Kuhn E, Lavielle M. Maximum likelihood estimation in nonlinear mixed effects models. Computational Statistics \& Data Analysis. 2005;49(4):1020-38.

30. Lindstrom M, Bates D. Nonlinear mixed effects models for repeated measures data. Biometrics. 2000;46(3):673–87.

31. Proust-Lima C, Philipps V, Liquet B. Estimation of extended mixed models using latent classes and latent processes: the R package lcmm. Journal of Statistical Software, 78(2), 1–56. 

32. Wickham H. ggplot2: elegant graphics for data analysis. Springer; 2016.

33. Jr CJ, Knopman D, Jagust W, Shaw L, Aisen P, Weiner M, et al. Hypothetical model of dynamic biomarkers of the Alzheimer’s pathological cascade. Lancet Neurol. 2010;9:119–28.

34. Buchman A, Capuano A, VanderHorst V, Wilson R, Oveisgharan S, Schneider J, et al. Brain beta-Amyloid Links the Association of Change in Body Mass Index With Cognitive Decline in Community-Dwelling Older Adults. J Gerontol A. 2023;78(2):277–85.

35. Wilson R, Capuano A, Bennett D, Schneider J, Boyle P. Temporal course of neurodegenerative effects on cognition in old age. Neuropsychology. 2016;30(5):591–9.

36. Wilson R, Capuano A, Sytsma J, Bennett D, Barnes L. Cognitive aging in older Black and White persons. Psychol Aging. 2015;30(2):279–85.

37. Wilson R, Capuano A, Marquez D, Amofa P, Barnes L, Bennett D. Change in Cognitive Abilities in Older Latinos. J Int Neuropsychol Soc. 2016;22(1):58–65.

38. Gerstorf D, Ram N, Mayraz G, Hidajat M, Lindenberger U, Wagner G, et al. Late-life decline in well-being across adulthood in Germany, the United Kingdom, and the United States: Something is seriously wrong at the end of life. Psychol Aging. 2010;25(2):477–85.

39. Terrera G, Minett T, Crayne C, Matthews F. Education associated with a delayed onset of terminal decline. Age Ageing. 2014;43(1):26–31.

40. Banesh D, Petersen M, Wendelberger J, Ahrens J, Hamann B. Comparison of piecewise linear change point detection with traditional analytical methods for ocean and climate data. Environ Earth Sci. 2019;78(21):1–16.

41. Ozekici S. Reliability and maintenance of complex systems. Springer Science \& Business Media; 2013.

\end{document}